\documentclass[showkeys,prl,preprint,showpacs,preprintnumbers,amsmath,amssymb]{revtex4}

\usepackage{graphicx}% Include figure files
\usepackage{dcolumn}% Align table columns on decimal point
\usepackage{bm}% bold math

\begin{document}

\preprint{}

\title{Euler-Lagrange correspondence of generalized Burgers cellular
automaton}% Force line breaks with \\

\author{Junta Matsukidaira and Katsuhiro Nishinari} 

\affiliation{
Department of Applied Mathematics and Informatics, 
Ryukoku University, Shiga 520-2194, Japan\\
}

\date{\today}% It is always \today, today,
             %  but any date may be explicitly specified

\begin{abstract}
 Recently, we have proposed a {\em Euler-Lagrange transformation} for
 cellular automata(CA) by developing new transformation
 formulas. Applying this method to the Burgers CA(BCA), we have
 succeeded in obtaining the Lagrange representation of the BCA. In this
 paper, we apply this method to multi-value generalized Burgers CA(GBCA)
 which include the Fukui-Ishibashi model and the quick-start model
 associated with traffic flow. As a result, we have succeeded in
 clarifying the Euler-Lagrange correspondence of these models. It turns
 out, moreover that the GBCA can naturally be considered as a simple
 model of a multi-lane traffic flow.
\end{abstract}

% PACS, the Physics and Astronomy
                             % Classification Scheme.
%Use showkeys class option if keyword
                              %display desired
\maketitle

\section{Introduction}

Cellular automata(CA) are simple objects that can describe not only
physical systems but also complex systems such as chemical, biological
and social systems. Since Wolfram has introduced the elementary CA (ECA)
in one-dimension as a model of complex system\cite{Wo}, lots of studies
have been done in these fields.

Among them, conservative CA\cite{HT,BF,F}, including soliton
systems\cite{TS,TM2} and traffic-flow models\cite{FI,NS}, have attracted
much attention because of their analytical properties which can be
treated by various mathematical methods.

In recent years, the authors have proposed and studied the so-called
ultra-discrete method by which we obtain corresponding CA from
difference equations\cite{TTMS,MSTTT,TM,TTM,NT,NMT}. Various soliton
equations have been transformed into CA by this method and the solutions
of the CA have also been obtained exactly without losing the
mathematical properties of corresponding difference equations. In this
method, max-function plays an important role through the ultra-discrete
formula,
\begin{equation}
 \lim_{\epsilon\rightarrow +0}\epsilon\log\left(\exp\left(\frac{A}{\epsilon}\right)+\exp\left(\frac{B}{\epsilon}\right)\right) = \max(A,B).
\end{equation}

On the other hand, it is known that several conservative CA allow two
different representation, {\it Euler} representation and {\it Lagrange}
representation\cite{Ni}. In the Euler representation, particles are
observed at a certain fixed point in space as a dependent(field)
variable, while in the Lagrange representation, we trace each particle
and follow the trajectory of it. Thus a dependent variable represents
the position of each particle in the Lagrange representation. In the
previous paper\cite{MN}, we have proposed a Euler-Lagrange
transformation for CA by developing new explicit transformation
formulas. We started from Burgers CA(BCA), which is the Euler
representation of rule-184 CA,
\begin{equation}  \label{BCA}
  U_j^{t+1} = U_j^t + \min(U_{j-1}^t, 1 - U_j^t)
                    - \min(U_j^t, 1 - U_{j+1}^t),
\end{equation}
where $U_j^t$ denote the number of particles at the site $j$ and time
$t$ and introducing the variable $S$ by
\begin{equation}
 S_j^t = \sum_{k=-\infty}^j U_k^t,
\end{equation}
or $U^t_j=S^t_j-S^t_{j-1}$,
and putting
\begin{equation} \label{EL}
 S_j^t = \sum_{i=0}^{N-1} H(j - x_i^t),
\end{equation}
where $H(x)$ is the step function defined by $H(x) = 1$ if $x \ge 0$ and
$H(x) = 0$ otherwise, we obtain Lagrange representation of
rule-184,
\begin{eqnarray}
 x_i^{t+1} &=& \min (x_i^t + 1, x_{i + 1}^t - 1) \nonumber\\
           &=& x_i^t + \min (1, x_{i + 1}^t - x_i^t - 1),
\end{eqnarray}
where $x^t_i$ is the Lagrange variable that represents the position of
the $i$-th particle at time $t$.

In deriving the Lagrange representation from Euler representation,
following two formulae for max function and step function play important
roles.

One is 
\begin{equation}\label{formula1}
 \sum_{k=1}^n H(j-\min(a_k,b_k))= \max\left(\sum_{k=1}^n H(j-a_k),\sum_{k=1}^n H(j-b_k)\right),
\end{equation}
where we assume that $a_1<a_2<\cdots<a_n$ and $b_1<b_2<\cdots<b_n$.
This formula expresses the commutability of max function and 
step function. Another formula is
\begin{equation}\label{formula2}
 \max\left(\sum_{i} H(j-a^t_i)-m,0\right)=\sum_{i} H(j-a^t_{i+m}),
\end{equation}
where we assume that $a^t_j=\infty$ if $j$ is larger than the number of
particles.

In the Euler-Lagrange transformation, max-function again plays an
important role as in the ultra-discrete method. In addition, the step
function significantly contributes the transformation as well.

\section{generalized Burgers cellular automaton}

In this paper, we present a new model which is a multi-value
(multi-lane) and multi-neighbor (high-speed, long-perspective) extension
of BCA. The equation in the Euler representation is expressed by
\begin{equation}\label{GBCA}
 U_j^{t+1} - U_j^t = \min \left(\sum_{k=0}^{V-1}U_{j-1-k}^t,\ \sum_{k=0}^{P-1}(L-U_{j+k}^t)\right) - \min \left(\sum_{k=0}^{V-1}U_{j-k}^t,\ \sum_{k=0}^{P-1}(L-U_{j+1+k}^t)\right),
\end{equation}
where $U_j^t\in\{ 0,\cdots,L\}$ denotes the number of particles at the site $j$ and time
$t$.  The parameter $L$ represents the maximum capacity of a cell, $V$
represents the maximum speed of particles and $P$ represents perspective
of particles, i.e., the maximum number of particles that a particle can
see in its front. We call this model generalized Burgers cellular automaton
(GBCA) in this paper. It will become clear later that GBCA include Euler
representation of Fukui-Ishibashi model\cite{FI} and quick-start
model\cite{FB,NT2}, which are both known as traffic models.

Let us derive the Lagrange representation of GBCA according to the
method which we have proposed in the previous paper\cite{MN}.
First we introduce the variable $S$ by
\begin{equation}
 S_j^t = \sum_{k=-\infty}^j U_k^t,
\end{equation}
or $U^t_j=S^t_j-S^t_{j-1}$ where $S_j^t$ is the total number of the
particles from $-\infty$ to $j$-th site, which is assumed to be
finite. Rewriting (\ref{GBCA}) in $S$, we obtain
\begin{equation}\label{Seq}
 S_j^{t+1} = \max (S_{j-V}^t, S_{j + P}^t - LP).
\end{equation}
Here, we put 
\begin{equation}\label{EL2}
 S_j^t = \sum_{i=0}^{N-1} H(j - x_i^t),
\end{equation}
where $H(x)$ is the step function and $N$ is the total number of
particles on the cells.  $x^t_i$ is the Lagrange variable that
represents the position of the $i$-th particle at time $t$ and the
relation $x_0^t<x_1^t<\cdots<x_{N-1}^t$ holds.

Using (\ref{EL2}) to replace $S$ in (\ref{Seq}) by $H$, we have
\begin{equation}\label{tmpGX}
 \sum_{i=0}^{N-1}H(j-x^{t+1}_i) = \max\left(\sum_{i=0}^{N-1}H(j-x^t_i-V), \sum_{i=0}^{N-1}H(j-x^t_i+P)-LP\right).
\end{equation}

By using (\ref{formula1}) and (\ref{formula2}), we have
\begin{eqnarray}
 \sum_i H(j-x^{t+1}_i) &=& \max\left(\sum_i H(j-x^t_i-V),\sum_i H(j-x^t_{i+LP}+P)\right) \nonumber \\
 &=& \sum_i H\left(j-\min(x^t_i+V,x^t_{i+LP}-P)\right),
\end{eqnarray}

Comparing both sides, we finally obtain
\begin{eqnarray}\label{Lagrange}
 x_i^{t+1} &=& \min (x_i^t + V, x_{i + LP}^t - P) \nonumber\\
           &=& x_i^t + \min (V, x_{i + LP}^t - x_i^t - P)
\end{eqnarray}
This is the Lagrange representation of GBCA. If we put $L=1, P=1$,
(\ref{Lagrange}) becomes the Lagrange representation of the
Fukui-Ishibashi model\cite{FI},
\begin{equation}
 x_i^{t+1} = x_i^t + \min (V, x_{i + 1}^t - x_i^t - 1).
\end{equation}
Therefore we know the Euler representation of Fukui-Ishibashi model is
\begin{equation}
 U_j^{t+1} - U_j^t = \min \left(\sum_{k=0}^{V-1}U_{j-1-k}^t,\ 1-U_j^t\right) - \min \left(\sum_{k=0}^{V-1}U_{j-k}^t,\ 1-U_{j+1}^t\right).
\end{equation}

If we put $L=1, V=1$, we obtain the Lagrange representation of the
quick-start model\cite{FB,NT2},
\begin{equation}
 x_i^{t+1} = x_i^t + \min (1, x_{i + P}^t - x_i^t - P).
\end{equation}
Corresponding Euler representation of the quick-start model is
\begin{equation}\label{QS}
 U_j^{t+1} - U_j^t = \min \left(U_{j-1}^t,\ \sum_{k=0}^{P-1}(1-U_{j+k}^t)\right) - \min \left(U_j^t,\ \sum_{k=0}^{P-1}(1-U_{j+1+k}^t)\right).
\end{equation}
The expression in the case of $P=2$ has already been obtained in
\cite{NT2}. Thus (\ref{QS}) is a generalization of the previous result
to the arbitrary $P$.

\section{multi-value (multi-lane) CA}
In the previous section, we show that (\ref{GBCA}) in the case of $L=1$
includes the Euler representation of the Fukui-Ishibashi model and the
quick-start model, and clarify the Euler-Lagrange correspondence of
these models.

In this section, we show that (\ref{GBCA}) in the case of $L>1$ can be
interpreted as an multi-value (multi-lane) extension of the
Fukui-Ishibashi model or the quick-start model.

Let us consider (\ref{GBCA}) in the case of $P=1$ for simplicity. The
Euler representation becomes
\begin{equation}\label{MFIE}
 U_j^{t+1} - U_j^t = \min \left(\sum_{k=0}^{V-1}U_{j-1-k}^t,\ L-U_j^t\right) - \min \left(\sum_{k=0}^{V-1}U_{j-k}^t,\ L-U_{j+1}^t\right),
\end{equation}
and the corresponding Lagrange representation becomes
\begin{equation}\label{MFIL}
 x_i^{t+1} = x_i^t + \min (V, x_{i + L}^t - x_i^t - 1).
\end{equation}
We show in the following that (\ref{MFIE}) is a $(L+1)$-valued CA and
(\ref{MFIL}) describes the movement of particle in $L$-lane, which means
(\ref{MFIE}) and (\ref{MFIL}) are considered as a $(L+1)$-valued
($L$-lane) extension of Fukui-Ishibashi model.

Assume that $L>0$ and $0\le U_j^t\le L$ for any $j$ at a certain
$t$. Then, relations
\begin{equation}
 \begin{array}{lll}
  \displaystyle \min \left(\sum_{k=0}^{V-1}U_{j-1-k}^t,\ L-U_j^t\right) & \ge & 0  \\
  \displaystyle \min \left(\sum_{k=0}^{V-1}U_{j-k}^t,\ L-U_{j+1}^t\right) & \ge & 0 \\
  \displaystyle \min \left(\sum_{k=0}^{V-1}U_{j-1-k}^t,\ L-U_j^t\right) + U_j^t & = & \displaystyle \min \left(\sum_{k=-1}^{V-1}U_{j-1-k}^t,\ L\right)\le L \\
  \displaystyle \min \left(\sum_{k=0}^{V-1}U_{j-k}^t,\ L-U_{j+1}^t\right) - U_j^t & = & \displaystyle \min \left(\sum_{k=1}^{V-1}U_{j-k}^t,\ L-U_j^t-U_{j+1}^t\right) \\
  & \le & \displaystyle \min \left(\sum_{k=0}^{V-1}U_{j-1-k}^t,\ L-U_j^t\right)
 \end{array}
\end{equation}
holds. Therefore, $0\le U_j^{t+1} \le L$ holds for any $j$. This means
the equation (\ref{MFIE}) under the above condition is equivalent to a CA
with a value set $\{0, 1, \ldots, L\}$. 

In the case of $L=2$ and $V=2$, the time evolution of (\ref{MFIE}) is
illustrated by the FIG. 1,
\begin{figure}[htbp]
 \begin{center}
  \includegraphics[scale=0.8]{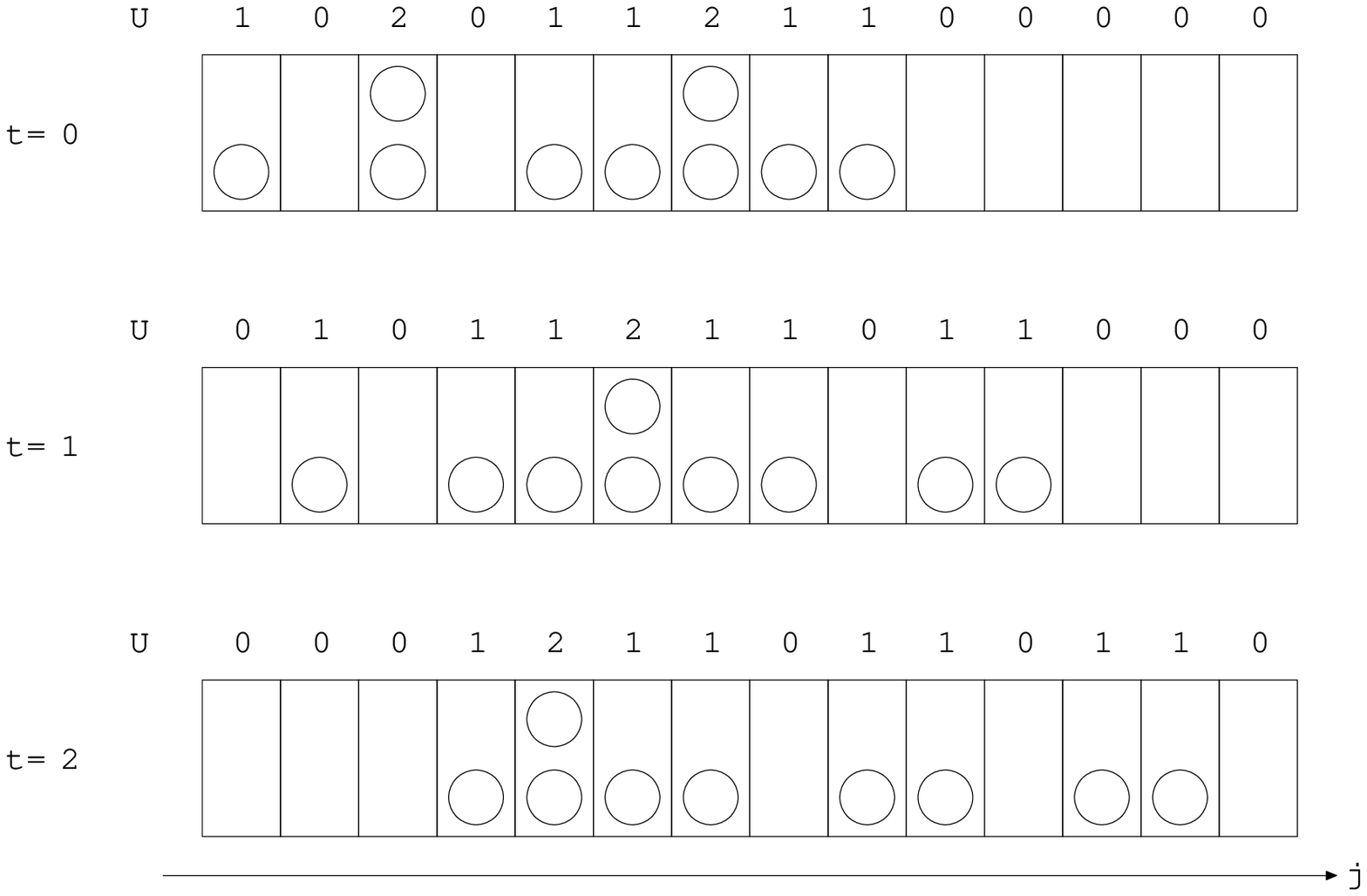} 
 \end{center}
 \caption{time evolution of the equation (\ref{MFIE})}
\end{figure}
where the number of the particles in the $j$th cell is equals to the
value of $U_j^t$. We can consider this case as a 3-valued CA in the
Euler representation.

Next, we introduce an ordering line on cells as shown in FIG. 2 . We number
particles along this line as shown in FIG.3, where $i$ denotes
suffix of variable $x_i^t$ in (\ref{MFIL}). For example, particles of
the initial condition in FIG. 1 are numbered as FIG. 4.

\begin{figure}[htbp]
 \begin{center}
  \includegraphics{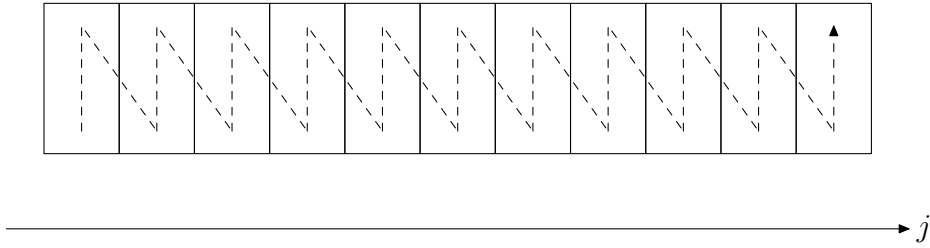} 
 \end{center}
 \caption{ordering line}
\end{figure}

\begin{figure}[htbp]
 \begin{center}
  \includegraphics[scale=1.0]{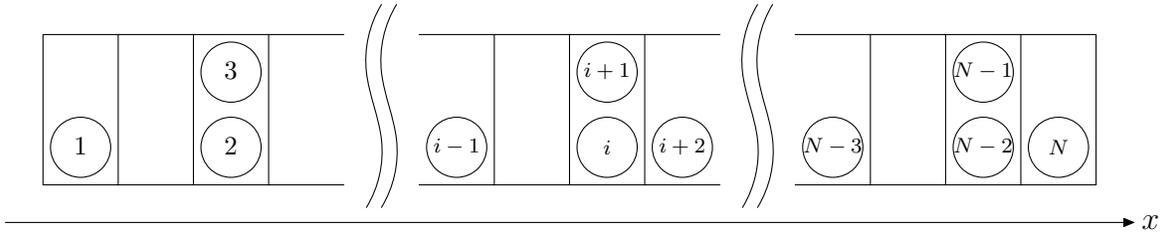} 
 \end{center}
 \caption{numbered particles along ordering line}
\end{figure}

\begin{figure}[htbp]
 \begin{center}
  \includegraphics[scale=0.8]{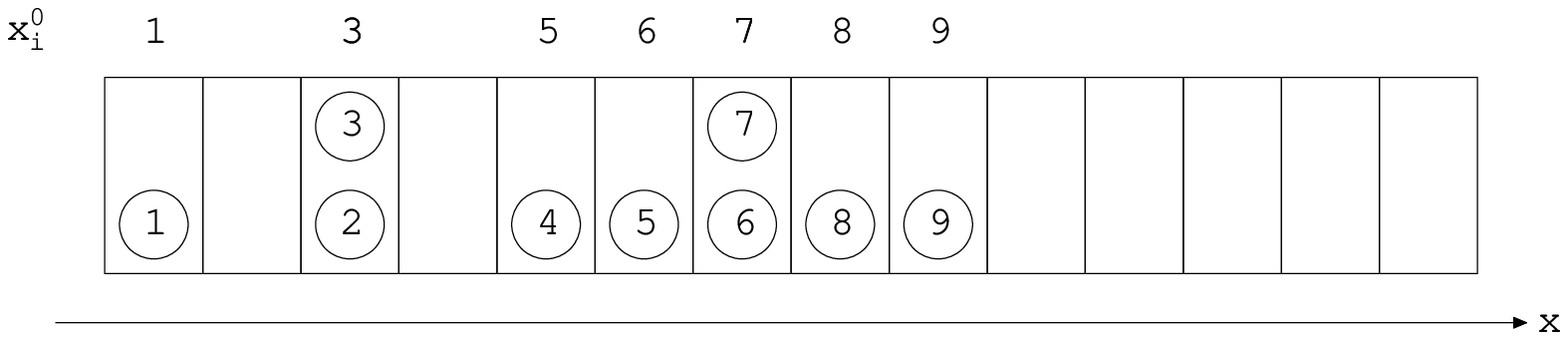} 
 \end{center}
 \caption{numbered particles at $t=0$ of FIG.1}
\end{figure}

If we take FIG. 4 as a initial condition for (\ref{MFIL}) and move
particles according to (\ref{MFIL}), we obtain FIG. 5. The
configurations of particles in FIG. 5 is identical with FIG. 1 if we
neglect numbers assigned to particles. Note here that although
(\ref{MFIL}) is a form of a car following model, the exclusion principle
in one cell, which means $x_i \ne x_j$ holds for any $i,j$ , does not
hold any more. Actually we see $x_4^1=x_5^1$ surely holds in FIG
5. Without the exclusion principle, (\ref{MFIL}) works successfully
together with (\ref{MFIE}), and describe the movement of the
particles. Note that there are no overtaking particles in the course of
time on the ordering line. As the the number of particle in one cell
remains less than 2, the model is considered as a car following model in
2-lane in this case.

\begin{figure}[htbp]
 \begin{center}
  \includegraphics[scale=0.8]{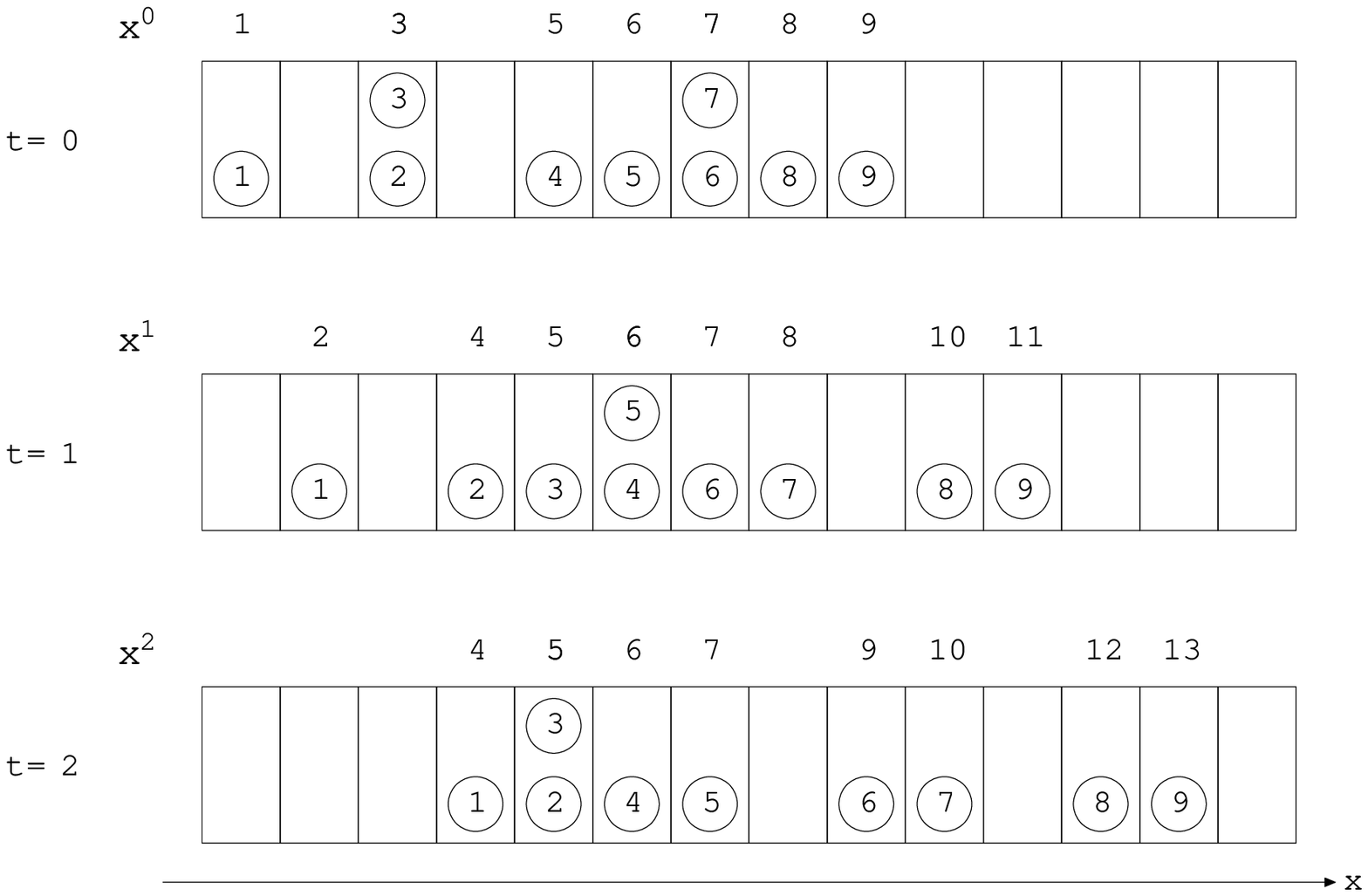} 
 \end{center}
 \caption{time evolution of the equation (\ref{MFIL})}
\end{figure}

Generally, as (\ref{MFIE}) is $(L+1)$-valued CA and (\ref{MFIL}) is a
car-following form of equation like the Fukui-Ishibashi model, we can
consider (\ref{MFIE}) and (\ref{MFIL}) as $(L+1)$-valued ($L$-lane)
extension of Fukui-Ishibashi model.

In the case of the quick-start model, we can similarly construct
multi-value (multi-lane) extension of the model although we don't
discuss in detail in this paper.

\section{Fundamental diagrams}

In this section, we discuss about fundamental diagrams of GBCA. In the
following, we consider a periodic boundary condition. GBCA can be
expressed in a conservation form such as
\begin{equation}
 \Delta_t U_j^t + \Delta_j q_j^t = 0,
\end{equation} 
where $\Delta_t$ and $\Delta_j$ are the forward difference operator with
respect to the indicated variable and $q_j^t$ represents a flow. Average
density $\rho$ and average flow $Q^t$ over all sites are defined by
\begin{equation}
 \rho \equiv \frac{1}{KL}\sum_{j=1}^K U_j^t, 
  \qquad Q^t \equiv \frac{1}{KL}\sum_{j=1}^K q_j^t
\end{equation}
where $K$ is the number of sites in a period. Since GBCA is in a
conservation form, average density does not depend on time and we can
use $\rho$ without a superscript $t$.

As we have shown, GBCA can be expressed by (\ref{Seq}). As (\ref{Seq})
can be transformed into ultra-discrete diffusion type equation,
asymptotic behavior of GBCA is similar to that of BCA. Since a pattern
of the maximum flow is given by $\cdots\underbrace{11\cdots
11}_P\underbrace{00\cdots 00}_V\cdots$, the density and the flow at the
phase transition from free to congested state become $\displaystyle
\frac{P}{P+V}$ and $\displaystyle \frac{PV}{P+V}$,
respectively\cite{Ni}. Thus the fundamental diagram is given by FIG. 6.

\begin{figure}[htbp]
 \begin{center}
  \includegraphics[scale=1.0]{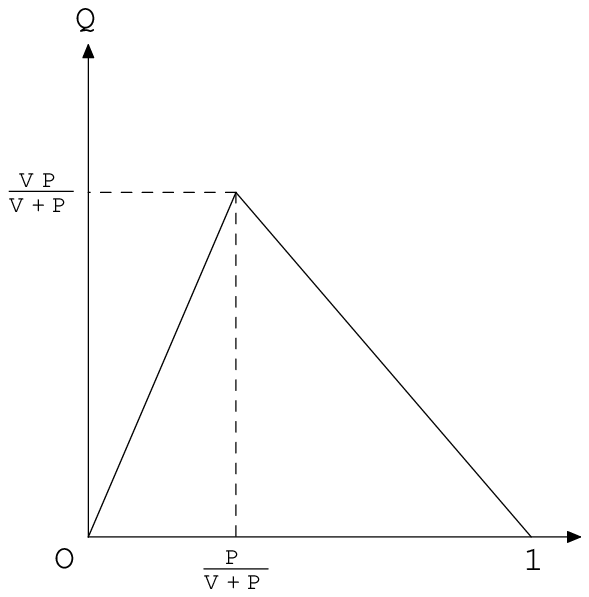} 
 \end{center}
 \caption{}
\end{figure}

\clearpage

\section{Concluding discussions}
In this paper, we have proposed GBCA which is a multi-value (multi-lane)
and multi-neighbor (high-speed, long-perspective) extension of BCA. We
have derived the Lagrange representation of GBCA, which includes the
Fukui-Ishibashi model and the quick-start model when $L=1$. By using the
Euler-Lagrange correspondence for GBCA, we have also obtained the Euler
representation of Fukui-Ishibashi model. Furthermore, we have obtained
multi-value (multi-lane) extension of the Fukui-Ishibashi model and the
quick-start model, and shown the multi-lane interpretation by
introducing ordering line on cells. It is worth mentioning again here
that the success of the Euler-Lagrange transformation in the multi-lane
case is due to the relaxation of the exclusion principle in
(\ref{MFIL}).

We have applied our method to more general cases than the case in the
previous papers and it works successfully. There are more several
conservative CA model which allows both Euler and Lagrange
representation. For examples, soliton systems called box and ball
system\cite{TS} are known to allow two different representation. Now we
are progressing on finding the Euler-Lagrange correspondence for soliton
systems and shall report them in the future.

We believe that establishing the general Euler-Lagrange correspondence
of CA will make a new development of the ultra-discrete method and
studies of CA. 
\section{Acknowledgments}

\begin{acknowledgments}
This work is supported in part by a Grant-in-Aid from the
 Japan Ministry of Education, Science and Culture.
\end{acknowledgments}

\newpage %Just because of unusual number of tables stacked at end
\bibliography{el2}% Produces the bibliography via BibTeX.
\end{document}